# Control of Electrochemical Corrosion Properties by Influencing Mn Partitioning through Intercritically Annealing of Medium-Mn Steel

René Daniel Pütz[1], Tarek Allam[2,3], Junmiao Wang[1], Jakub Nowak[1], Christian Haase[2], Stefanie Sandlöbes-Haut[4], Ulrich Krupp[2] and Daniela Zander[1,*]

[1] Chair of Corrosion and Corrosion Protection, Foundry Institute, RWTH Aachen University, Intzestr. 5, 52072 Aachen, Germany; r.puetz@gi.rwth-aachen.de (R.D.P.); j.wang@gi.rwth-aachen.de (J.W.); j.nowak@gi.rwth-aachen.de (J.N.); d.zander@gi.rwth-aachen.de (D.Z.)

[2] Chair Materials Engineering of Metals, Steel Institute, RWTH Aachen University, Intzestr. 1, 52072 Aachen, Germany; tarek.allam@iehk.rwth-aachen.de (T.A.); christian.haase@iehk.rwth-aachen.de (C.H.); krupp@iehk.rwth-aachen.de (U.K.)

[3] Department of Metallurgical and Materials Engineering, Suez University, 43528 Suez, Egypt (T.A.)

[4] Institute for Physical Metallurgy and Materials Physics, RWTH Aachen University, Kopernikusstr. 14, 52074 Aachen, Germany; sandloebes-haut@imm.rwth-aachen.de (S.S.)

* Correspondence: Daniela Zander, d.zander@gi.rwth-aachen.de (D.Z.)

**Abstract:** Medium-Mn steels exhibit excellent mechanical properties and lower production costs compared to high-Mn steels, which makes them a potential material for future application in the automotive industry. Intercritical annealing (ICA) after cold rolling allows to control the stacking fault energy (SFE) of austenite, the fraction of ferrite and reverted austenite, and the element partitioning (especially Mn). Although Mn deteriorates the corrosion behavior of Fe-Mn-Al alloys, the influence of austenite fraction and element partitioning of Mn on the electrochemical corrosion behavior has not been investigated yet. Therefore, the electrochemical corrosion behavior in 0.1 M $H_2SO_4$ of X6MnAl12-3, which was intercritically annealed for 2 h at 550 °C, 600 °C and 700 °C, was investigated by potentiodynamic polarization (PDP), electrochemical impedance spectroscopy (EIS) and mass spectroscopy with inductively coupled plasma (ICP-MS). Additionally, specimens after 1 h and 24 h of immersion were examined via SEM to visualize the corrosion damage. The ICA specimens showed a selective dissolution of reverted austenite due to its micro-galvanic coupling with the adjacent ferrite. The severity of the micro-galvanic coupling can be reduced by decreasing the interface area as well as the chemical gradient of mainly Mn between ferrite and reverted austenite by ICA.

## 1. Introduction

Advanced high strength steels (AHSS) have been developed since the 1980s to fulfill the low emission and high fuel-efficiency demands in the automotive industry [1]. Significant research has been conducted in the development and application of AHSS throughout the world. Compared to conventional high-strength steels, AHSS strongly reduce vehicle weight and show a better combination of strength, ductility, and toughness, which enhances crash performance and vehicle safety. Recently, the third generation AHSS is of great interest because of the combination of strength, formability and lower production costs in comparison to the first and second generation AHSS [2-4]. Medium-Mn steels belong to the third generation of AHSS, which show excellent me-chanical properties at reduced Mn content and production costs compared to high-Mn steels [5,6].

Medium-Mn steels have an ultrafine grained duplex microstructure containing ferrite (α-Fe) and reverted austenite (γ-Fe), resulting from inter-critical annealing (ICA) treatments in the α + γ region. In addition to Mn, Al and/or Si are important alloying elements in alloy design of medium-Mn steels, since they play a crucial role in suppressing cementite formation, expand the ICA temperature-window and control the stacking fault energy (SFE) of austenite [7]. Another important aspect during ICA treatments is the local element partitioning that is responsible for the stability of reverted austenite and accordingly tuning the mechanical properties [8-10]. Many researchers have investigated the influence of ICA on the microstructure and mechanical properties of medium-Mn steels. It was found that the ICA temperature strongly affects the volume fraction, morphology and stability of reverted γ-Fe [3,5,11-14]. Reverted γ-Fe transforms into α´-martensite and/or twins during deformation, influencing the strain-hardening behavior and ductility. It is reported that the ICA time has effects on the morphology of reverted γ-Fe and the mechanical properties [5,15]. Besides, the initial microstructure also plays a role in the phase morphologies in medium-Mn steels [14,16-19].

Since corrosion is a notorious issue for steels, the corrosion behavior of Mn steels is of high interest for potential application. Only some investigation of the corrosion behavior of high-Mn steels has been conducted. It is found that high Mn content adversely affected the corrosion resistance of high-Mn steels in various electrolytes like 1 M $Na_2SO_4$ [20] and 30 % NaOH [21] as well as 50 % NaOH [22] due to the instable Mn-rich oxides formed in the corrosion products. Additionally, a high-Mn steel showed also a detrimental corrosion resistance in 0.1 M $H_2SO_4$ compared to an interstitial-free steel due to the high dissolution tendency of Mn and Fe in the acidic solution [23]. Additions of alloy elements in high-Mn steels could enhance the corrosion resistance due to the beneficial effects on passivation, such as Cr in 3.5 wt.% NaCl [24,25], 10–50% $HNO_3$ [22], and 1 M $Na_2SO_4$ solution [20,22], Al in 1 M [21] and 5 wt.% $Na_2SO_4$ [26], 10-50% NaOH [21], 3.5 wt.% NaCl [24], 1 N $H_2SO_4$ [27] and 50% $HNO_3$ [20] and Si in 1 N $H_2SO_4$ [27]. Meanwhile, it was observed that γ-Fe was more resistant to corrosion compared to α-Fe in 3.5 wt.% NaCl solution [24,28]. The positive influence of grain refinement on the corrosion behavior of high-Mn steels in 3.5 wt.% was also reported [29]. With growing interest in medium-Mn steels, the corrosion behavior of medium-Mn draws attention as well. Several authors [30-32] also reported adverse effects of Mn on the corrosion behavior of medium-Mn steel containing 5.5 wt.% Mn in simulated seawater and 3.5 wt.% as well as 5 wt.% NaCl solution, which is similar to high-Mn steels. Besides, it was observed that the addition of alloying elements like Cr, Mo, Cu and Ni improved the corrosion behavior in simulated seawater and 5.0 wt.% NaCl solution [30,32]. Allam et al. [6] developed an austenitic medium-Mn steel with enhanced corrosion resistance in 5 wt.% NaCl solution due to a higher Cr content, ultrafine microstructure and the addition of N. However, more research is needed to comprehensively clarify the corrosion behavior of medium-Mn steels. Especially the influence of several heat treatment conditions on the corrosion behavior of medium-Mn steels is hardly existent.

In the present study, a medium-Mn steel X6MnAl12-3 in 0.1 M $H_2SO_4$ was investigated in order to evaluate the influence of intercritical annealing and the concomitant change in microstructure on the electrochemical corrosion behavior focusing on the interaction between ferrite and reverted austenite.

## 2. Materials and Methods

### 2.1. Materials processing

The chemical composition of the medium-Mn steel X6MnAl12-3 under investigation is listed in Table 1. It was melted in a lab-scale vacuum induction furnace to produce a cast-ingot of 140 x 140 x 550 mm³. The cast-ingot underwent hot forging at 1150 °C to a thickness of 40 mm followed by hot rolling at 1150 °C to manufacture hot-rolled sheets with a thickness of 3 mm. These hot-rolled sheets were subjected to a homogenization heat treatment at 1100 °C for two hours and a subsequent water quench. Cold rolling processes were conducted at room temperature with a total thickness reduction of 50 % to prepare cold-rolled (CR) sheets of 1.5 mm in thickness. For the subsequent electro-chemical investigations, rectangular specimens (11 mm x 10 mm x 1.5 mm) were segmented from the CR state by means of wire-cut electrical discharge machining (EDM). The target microstructures of the prepared specimens containing different volume fractions of austenite phase were adjusted through inter-critical annealing (ICA) treatments in a salt bath at different temperatures, namely; 550, 600 and 700 °C for two hours followed by a water quench. The ICA treatments were designed based on thermodynamic calculations performed using Thermo-Calc® software (Thermo-Calc Software, Stockholm, SE) with the database TCFE10 for Steels/Fe-alloys. A schematic illustration of the laboratory processing route of the investigated steel X6MnAl12-3 is shown in Figure 1.

**Table 1.** Chemical composition in wt.% of the X6MnAl12-3 medium-Mn steel under investigation.

| Steel grade | C | Si | Mn | P | S | Al |
|---|---|---|---|---|---|---|
| X6MnAl12-3 | 0.064 | 0.2 | 11.7 | 0.006 | 0.003 | 2.9 |

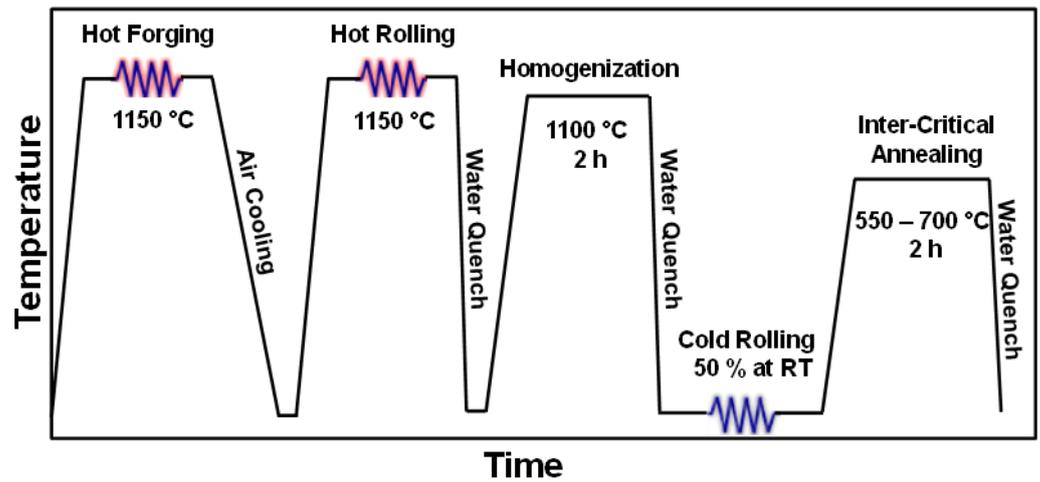

**Figure 1.** Schematic illustration of the laboratory processing route of the investigated steel X6MnAl12-3.

### 2.2. Characterization of the microstructures

The microstructures after cold rolling and annealing were investigated using a Zeiss Gemini field-emission gun scanning electron microscope (FEG-SEM) (Carl Zeiss Microscopy GmbH, Jena, DE) equipped with an additional Oxford X-Max50 energy dispersive X-ray spectrometer (EDS) (Oxford Instruments, Abingdon, UK). The applied acceleration voltages and working distances during SEM investigations were 15-20 kV and 10-20 mm, respectively. The CR and ICA specimens were prepared following the standard procedures for preparation of metallographic specimens starting with mechanical grinding up to grit 2400, followed by mechanical polishing using 6 and 1 μm diamond polishing pastes. Afterwards, the polished specimens were subjected to 3% Nital etching solution to reveal the features of the microstructures. The present phases and their volume fractions in different states (CR and ICA) were identified and quantified by means of X-ray diffraction (XRD) technique using a Bruker D8 advanced X-ray diffractometer (Bruker, Billerica, MA, US). Fe Kα radiation (0.639 nm) was used to acquire the diffractograms.

## 2.3. Electrochemical testing and corrosion analysis

The rectangular specimens were mounted in a non-conductive polymer resin with an exposed surface (RD/TD plane) area of approximately 1.1 cm². The specimens were stepwise ground and polished till 0.25 μm. Afterwards, the specimens were rinsed with distilled water, ethanol and then air-dried. The thin specimens were conductively glued with a Cu-cylinder from the bottom. A drilled hole acts as a connecting path for an isolated Cu-cable to the specimens. A glass tube was glued to the resin above the Cu-cable preventing the contact of the solution with the Cu-cable and the specimen during the electrochemical measurements. The samples were stored in a desiccator after gluing. A similar working electrode (WE) setup is described elsewhere [33].

The utilized electrolyte was 0.1 M $H_2SO_4$ with a pH of 1.1 ± 0.1. The electrolyte was prepared with analytic grade chemicals and bi-distilled water. Figure 2 shows the schematic illustration of the three-electrode setup for the electrochemical testing. A thermostat was used to maintain a constant electrolyte temperature of 25 °C and a Reference 600 potentiostat (Gamry Instruments Inc, Warminster, PA, US) was utilized to perform potentiodynamic polarization (PDP) and electrochemical impedance spectroscopy (EIS) measurements. The specimen was used as the working electrode. A platinum sheet, which exhibits a larger surface area compared with the working electrode, was used as the counter electrode (CE). A saturated calomel electrode (SCE), which was connected via a Haber-Luggin capillary with the electrolyte in the beaker and placed approximately 1 cm in front of the specimen, was used as reference electrode (RE). A solution volume of 700 ml was used for PDP and EIS experiments. The solutions were purged using Ar for 30 min prior to the open circuit potential (OCP) measurements of 60 min.

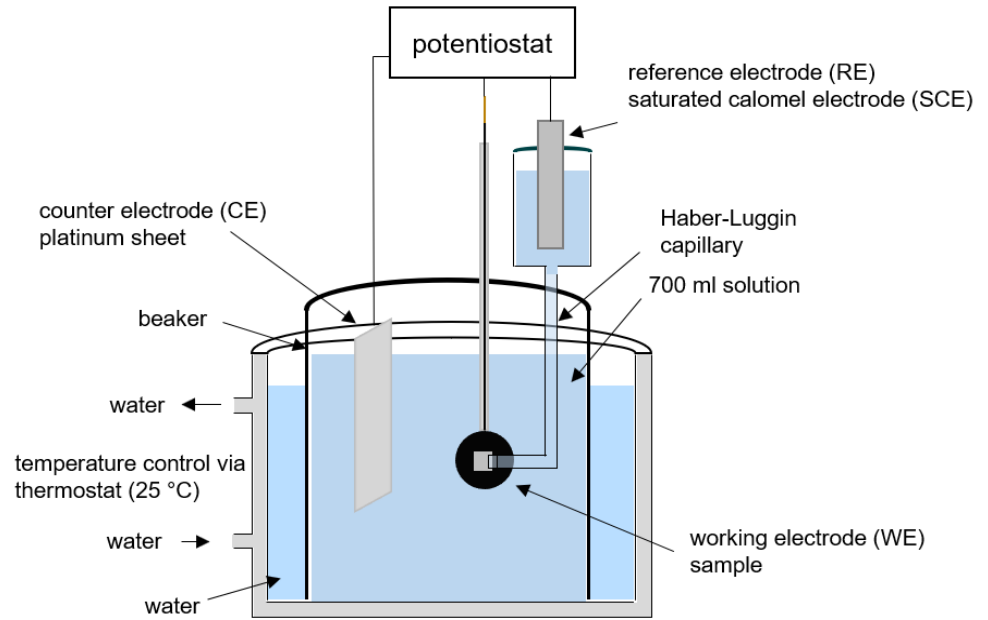

**Figure 2.** Experimental setup (three-electrode cell) for the electrochemical testing adapted from [33,34].

The PDP measurements were conducted in 0.1 M $H_2SO_4$ using a scan rate of 0.167 mV/s from -0.3 $V_{SCE}$ vs. OCP up to 0.25 $V_{SCE}$ vs. OCP. For all polarization tests, iR compensation was applied. The corrosion current densities ($i_{corr}$) were determined by graphical Tafel analysis using the anodic and cathodic branches.

Immersion tests with potentiostatic EIS measurements were utilized in 0.1 M $H_2SO_4$ at 25 °C for 24 h. The EIS measurements were performed at the OCP over a frequency range from 10 kHz to 0.1 Hz after 1, 4, 8 and 24 h, using 10 points per decade and applying an AC amplitude of 5 mV rms. Separate immersion tests without EIS measurements were performed in 0.1 M $H_2SO_4$ at 25 °C for 24 h. A solution sample of 5 ml was taken at specific time intervals, namely, after 0, 1, 4, 8 and 24 h. The chemical concentrations ([57]Fe, [55]Mn and [27]Al isotopes) of the solution samples were analyzed without dilution or additional acidification by NexION® 2000 mass spectroscopy with inductively coupled plasma (ICP-MS) (Perkin Elmer, Waltham, MA, US) in kinetic energy discrimination (KED) mode. The KED mode was used to diminish polyatomic ion interferences [35]. The purity of the used nebulizer and collision gas was Ar (99.999 vol.%) and He (99.9999 vol.%) with 0 % relative humidity. Three solution samples were investigated for each heat treatment condition. Top view and cross-section images of the rim zone after the immersion tests were performed using a Supra 55 VP scanning electron microscope (SEM) (Carl Zeiss Microscopy GmbH, Jena, DE).

## 3. Results

### 3.1. Thermodynamic caculations

The equilibrium austenite and ferrite phase fractions as well as their corresponding chemical compositions were predicted based on the thermodynamic calculations represented in Figure 3. Accordingly, the ICA treatments were designed to adjust the target microstructures with low, intermediate and high austenite fractions within the selected ICA range starting from 550, 600 to 700 °C as indicated in Figure 3. The evolution of phase fractions shows an obvious increase in equilibrium austenite fraction at the expense of ferrite phase within the ICA range from 0.42 at 550 °C, 0.52 at 600 °C to 0.78 at 700 °C. Such increase in austenite fraction with increasing the temperature within the ICA range is accompanied with elemental partitioning phenomena that control the subsequent microstructural features and the related electrochemical behaviors. It is clear that Mn and C, strong austenite stabilizers, generally tend to be enriched in austenite Figure 3 (a) as compared to ferrite Figure 3 (b). However, Al and Si, ferrite stabilizers, are more enriched in ferrite than in austenite. As the ICA temperature increases, the Mn content in austenite decreases from almost 22 wt.% at 550 °C to around 18 wt.% at 600 °C and 13 wt.% at 700 °C. Moreover, a concurrent increase of the Al content in austenite is observed with increasing the ICA temperature reaching more than 2.2 and 2.6 wt.% at 600 and 700 °C, respectively, while it maintains below 2 wt.% at 550 °C. It is worth mentioning that compared with austenite, the Mn content in ferrite generally shows relatively low values below 5 wt.% at 550 and 600 °C and slightly higher than 5 wt.% at 700 °C. Meanwhile, the Al content in ferrite hardly changes with increasing ICA temperature and shows always higher values (around 4.5 wt.%) compared with its values in austenite phase. Likewise, the Si content in ferrite stays at a relatively high level of around 1 wt.% compared with its level in austenite that shows a slight decrease from around 0.36 to 0.23 wt.% when the ICA temperature increases from 550 to 700 °C, respectively.

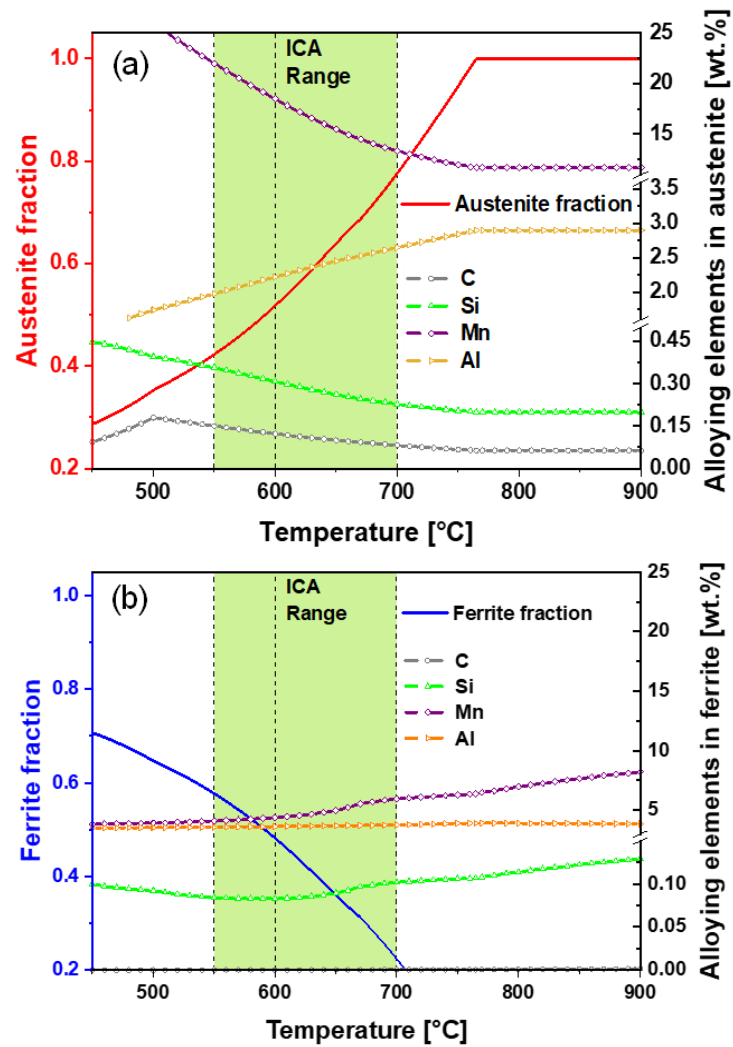

**Figure 3.** Thermodynamic calculations performed using Thermo-Calc® software package (TCFE10 for Steels/Fe-alloys). The evolution of phase fractions and their corresponding contents of alloying elements under thermodynamic equilibrium conditions are represented (a) for austenite and (b) ferrite.

### 3.2 Microstructure evolution during ICA treatment

The SEM micrographs of medium-Mn steel X6MnAl12-3 in CR as well as in ICA states are displayed in Figure 4. The CR state exhibits a fully martensitic microstructure with deformed martensite being preferentially elongated parallel to the rolling direction (Figure 4 (a)). However, the ICA states generally show dual-phase microstructures containing reverted austenite and ferrite grains. The relative amounts of the phases show an obvious dependency on the applied ICA temperatures. Figure 4 (b) shows the developed dual-phase microstructure after ICA at 550 °C followed by a water quench. The inset of Figure 4 (b) reveals the formation of ultrafine grained islands of austenite with a relatively small amount embedded in the ferrite matrix. Both the amount and size of austenite phase increase as the applied ICA temperature increases to 600 °C, as represented in Figure 4 (c). Furthermore, it can be observed from the inset of Figure 4 (c) that the austenite phase shows an elongated grain morphology, which is inherited from the deformed martensite grains. Such elongated morphology of austenite grains becomes less pronounced and changes to a mixed type of equiaxed and elongated morphologies as the applied ICA temperature increases further to 700 °C (inset of Figure 4 (d)). Moreover, the ICA treatment at 700 °C resulted in a clear increase in the amount of austenite phase. The XRD-measurements carried out to quantify the constituting phases of the CR and ICA states are shown in Figure 5. The XRD spectra depicted in Figure 5 (a) demonstrate the presence of the characteristic peaks of austenite phase for the ICA states (550, 600 and 700 °C), while the characteristic peaks of only ferrite/martensite appear for the CR state. The volume fractions of austenite phase in the ICA states are displayed as a bar chart analysis in Figure 5 (b) indicating the increase of the austenite volume fraction from 21.9 vol.% for ICA at 550 °C to 40.8 and 58.5 vol.% as the ICA temperature increases to 600 and 700 °C, respectively.

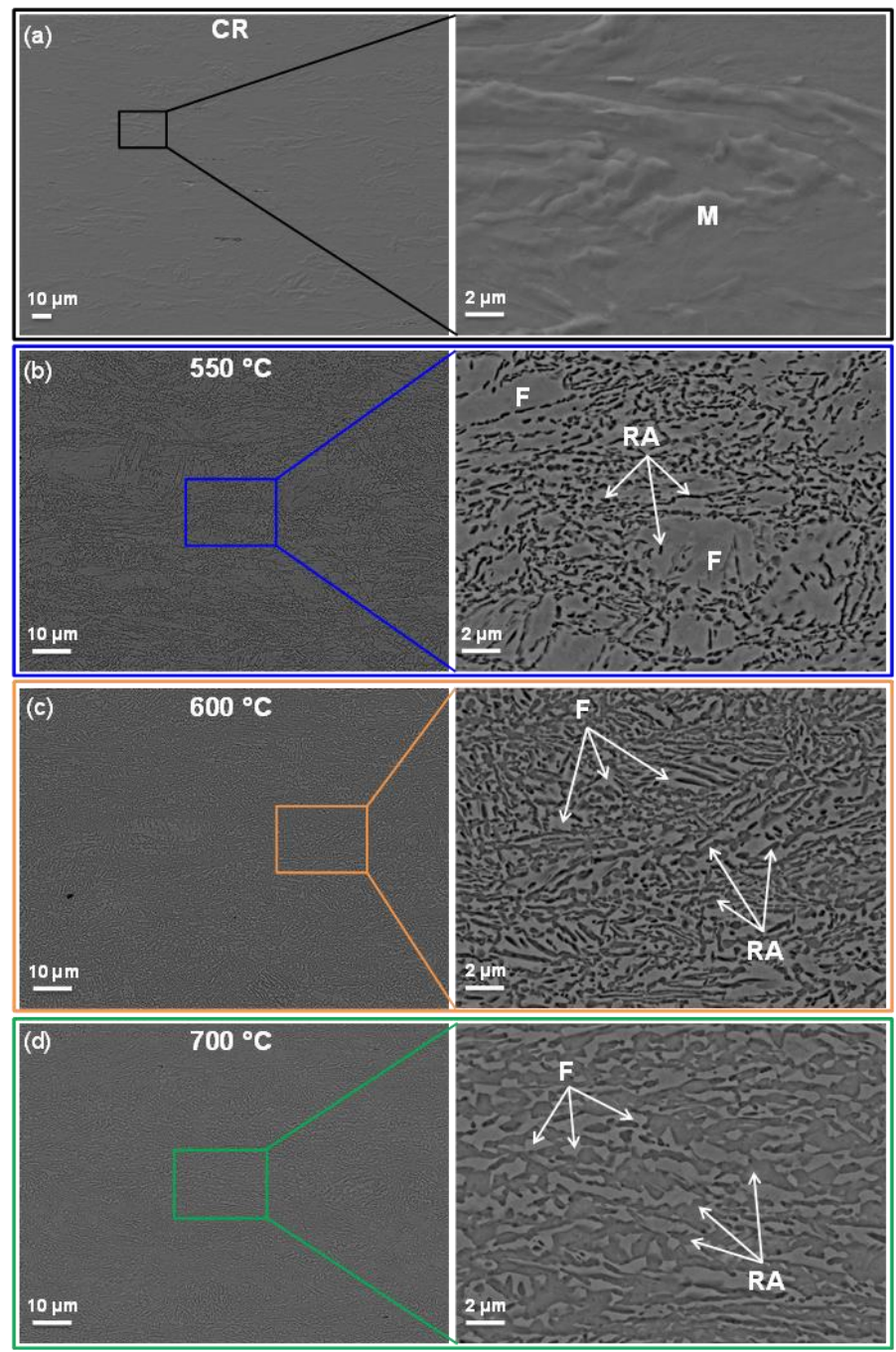

**Figure 4.** SEM micrographs of the investigated medium-Mn steel X6MnAl12-3 in different ICA states. (a) cold-rolled (CR) state. (b), (c) and (d) are for the ICA states annealed at 550, 600 and 700 °C, respectively. M, RA and F denote martensite, reverted austenite and ferrite, respectively.

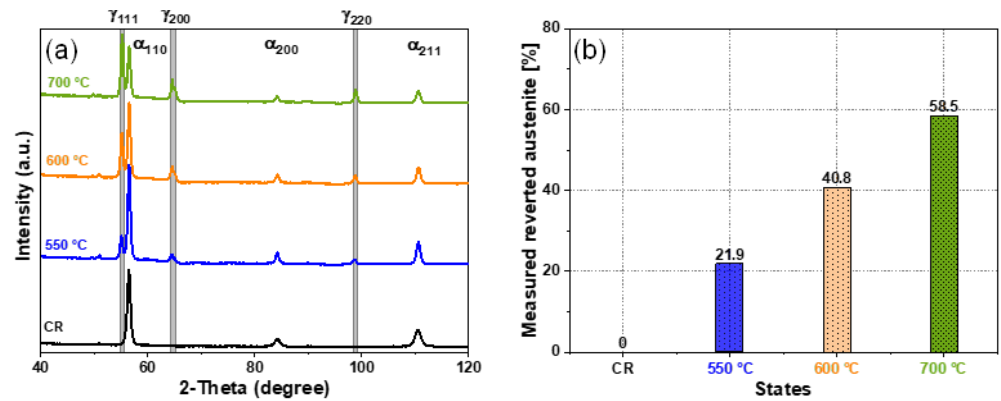

**Figure 5.** XRD measurements of the medium-Mn steel X6MnAl12-3 in different ICA states. (a) the recorded XRD spectra, (b) bar chart of the measured volume fraction of the reverted austenite phase.

*3.3 Electrochemical testing and corrosion analysis*

In Figure 6 (a), the potentiodynamic polarization curves of the medium-Mn steels in 0.1 M $H_2SO_4$ are shown. In general, all investigated medium-Mn steels revealed an active dissolution independently from the processing parameters. In Figure 6 (b), a bar graph of the determined corrosion current densities ($i_{corr}$) and the corrosion potentials ($E_{corr}$) is depicted. $E_{corr}$ of CR (-567 ± 2 mV$_{SCE}$) and ICA 550 °C (-567 ± 4 mV$_{SCE}$) are similar, while $E_{corr}$ of ICA 600 °C (-570 ± 1 mV$_{SCE}$) and ICA 700 °C (-575 ± 2 mV$_{SCE}$) decrease with increasing ICA temperature. The (CR) state shows the lowest current density value compared to the medium-Mn steels after intercritical annealing. In respect to the error bars, the order of the $i_{corr}$ values is CR < ICA 550 °C and ICA 700 °C < ICA 600 °C. The $i_{corr}$ values increase from the CR condition (0.70 ± 0.18 mA/cm²) to ICA 550 °C (1.19 ± 0.32 mA/cm²) and further to ICA 600 °C (2.69 ± 0.12 mA/cm²), then decrease to ICA 700 °C (1.23 ± 0.24 mA/cm²). The $i_{corr}$ value of ICA 700 °C (1.23 ± 0.24 mA/cm²) is comparable to that of ICA 550 °C (1.19 ± 0.32 mA/cm²). The determined electrochemical parameters including the corrosion current density ($i_{corr}$), anodic Tafel slope ($\beta_a$), cathodic Tafel slope ($\beta_c$) and corrosion potential ($E_{corr}$) of the potentiodynamic polarization in 0.1 M $H_2SO_4$ of all material states are listed in Table 2.

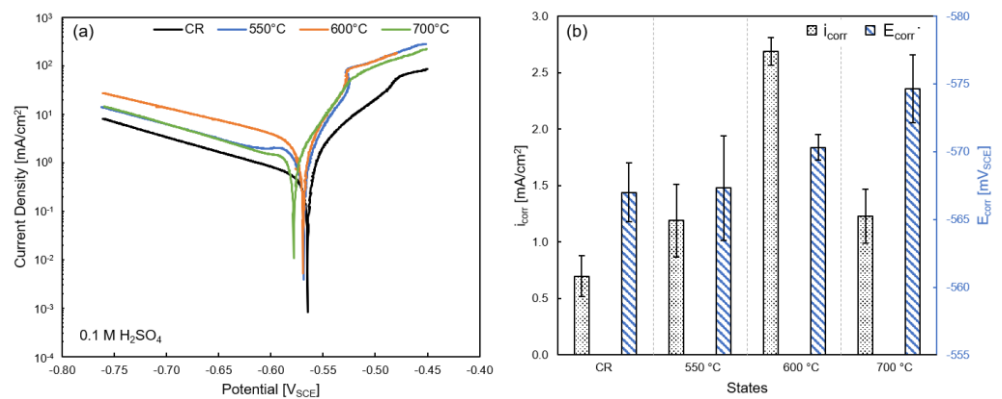

**Figure 6.** (a) Potentiodynamic polarization curves of the different medium-Mn states in 0.1 M $H_2SO_4$ and (b) bar graph of the $i_{corr}$ and $E_{corr}$ values.

**Table 2.** Electrochemical parameters in 0.1 M $H_2SO_4$.

| State | $i_{corr}$ [mA/cm²] | $\beta_c$ [V/decade] | $\beta_a$ [V/decade] | $E_{corr}$ [mV$_{SCE}$] |
|-------|--------------------|---------------------|---------------------|------------------------|
| CR | 0.70 ± 0.18 | -0.159 ± 0.011 | 0.034 ± 0.006 | -567 ± 2 |
| 550 °C | 1.19 ± 0.32 | -0.167 ± 0.000 | 0.029 ± 0.001 | -567 ± 4 |
| 600 °C | 2.69 ± 0.12 | -0.200 ± 0.000 | 0.038 ± 0.001 | -570 ± 1 |
| 700 °C | 1.23 ± 0.24 | -0.159 ± 0.011 | 0.034 ± 0.002 | -575 ± 2 |

The EIS Nyquist plots of all states at the open circuit potential in 0.1 M H$_2$SO$_4$ after an immersion time of 1 h are shown in Figure 7 (a). Each state contains two various time constants, which are represented by a capacitance semicircle at the mid-section of the frequency range and a semicircle in the low-frequency range. The mid-frequency circle is related to an electrochemical double layer, while the low-frequency circle reveals the existence of a pseudo-inductance, which is related to an adsorption process. The capacitive arc at the mid-frequency region of the different states show various diameters indicating deviating charge transfers and consequently different corrosion resistances. The CR state shows the largest circle, while the ICA treatments lead to a decrease of the circle diameters. ICA 700 °C shows the largest diameter and ICA 600 °C the lowest, while ICA 550 °C is located in between.

In Figure 7 (b), the Bode plots (phase angle and total impedance magnitude) of all states in 0.1 M H$_2$SO$_4$ at 1 h are depicted. The chronology of the total impedance magnitude of the four states correlate with the diameters in the Nyquist plots indicating an increase of the corrosion resistance from ICA 550 °C, to ICA 600 °C, to ICA 700 °C and finally up to the CR state. The phase angle Bode plot indicates the presence of two time constants, which correspond to the capacitive arc in the mid-frequency region and the pseudo-inductance (adsorption) in the low-frequency region.

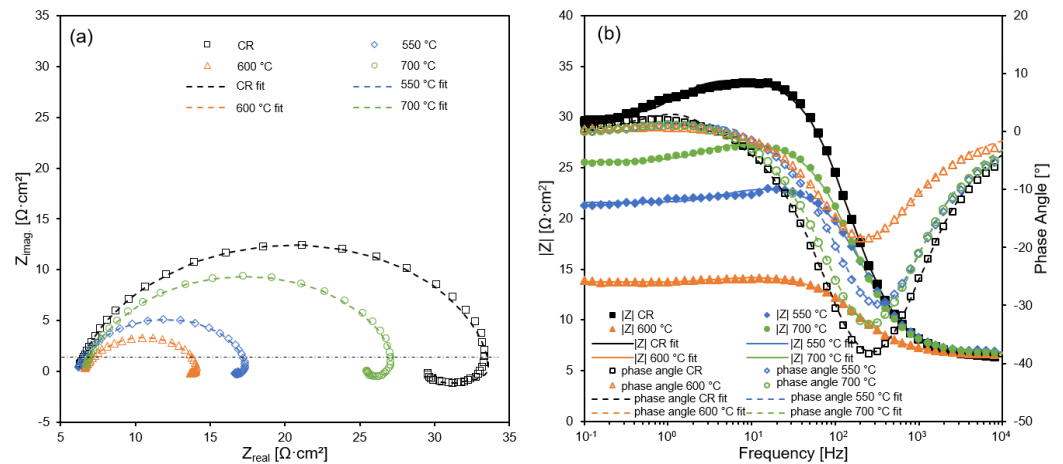

**Figure 7.** (a) Nyquist and (b) Bode plot of the specimens in all heat treatment states after 1 h immersion in 0.1 M H$_2$SO$_4$.

In order to fit the obtained EIS data, the equivalent electrochemical circuit model (EEC) in Figure 8 is used. The mid-frequency time constant is modelled with a Randles circuit, which is associated to an electrochemical double layer. The Randles circuit (Figure 8 (a)) contains the ohmic resistance of the solution ($R_{sol}$), the constant phase element of the double layer (CPE$_{DL}$), the charge transfer resistance ($R_{CT}$) and a general element $Z_{ads}$, which is placed in series with $R_{CT}$. $Z_{ads}$ represents the singular adsorbate model of the low-frequency time constant (pseudo-inductance) using the constant phase element of the adsorbates (CPE$_{ads}$), the adsorbate resistance ($R_{ads}$) and the adsorbate conductor ($L_{ads}$), which are arranged in parallel (Figure 8 (b)). The usage of CPE$_{DL}$ and $R_{CT}$ in series with $R_{sol}$ were also used in the literature for various Fe-based materials like steels [36,37] or iron aluminides [38,39] in H$_2$SO$_4$ containing solutions and are well suitable to fit the data in the present study.

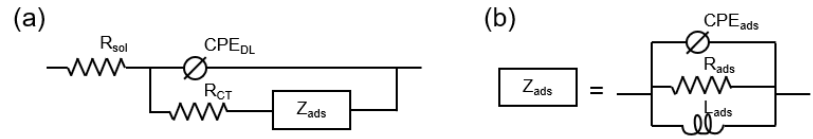

**Figure 8.** Electric equivalent circuit (EEC) containing (a) a Randles circuit used to evaluate the EIS data. (b) The parallel-connected general adsorbate element $Z_{ads}$.

In Figure 9, the charge transfer resistances over time of all states during immersion in 0.1 M $H_2SO_4$ are shown. The chronology of $R_{CT}$ is in line with the diameter of the Nyquist plot and the total impedance magnitude in the Bode plot confirming the decreasing corrosion resistance from CR, to ICA 700 °C, to ICA 550 °C, to ICA 600 °C after 1 h of immersion. After 4 h of immersion, a partial change in $R_{CT}$ is observed. The charge transfer resistance of CR and ICA 700 °C as well as ICA 550 °C and ICA 600 °C align after 4 h in the 0.1 M $H_2SO_4$ solution, respectively. After 24 h of immersion, ICA 700 °C shows the highest charge transfer resistance, while $R_{CT}$ of CR decease further and sort itself in between ICA 700 °C and ICA 550 °C. $R_{CT}$ of ICA 600 °C is similar with $R_{CT}$ of ICA 550 °C.

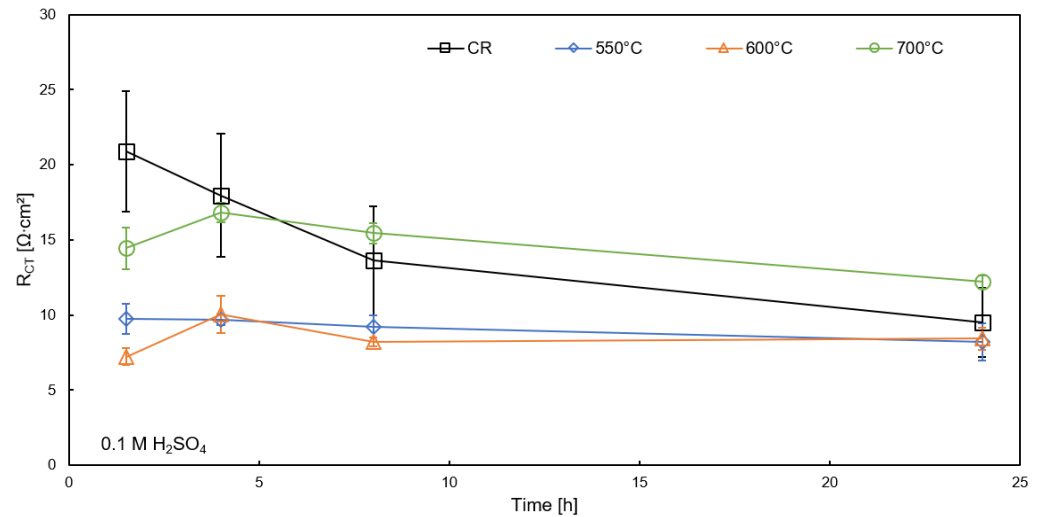

**Figure 9.** EIS resistance of charge transfer ($R_{CT}$) of the different heat treatment states at various time points (1, 4, 8 and 24 h) during immersion in 0.1 M $H_2SO_4$.

The EIS fitting parameters $R_S$, $R_{CT}$, $R_{ads}$, $L_{ads}$, the dimensionless fraction exponents of the double layer $n_{DL}$ and adsorbate $n_{ads}$, the admittance of the double layer $Y_{DL}$ and adsorbate $Y_{ads}$, and the fitting variance $X^2$ are listed in Table 3.

**Table 3.** Fitting parameters using EIS data based on the EEC in Figure 8 for various ICA states in 0.1 M $H_2SO_4$.

| Time [h] | $R_S$ [$\Omega \cdot cm^2$] | $R_{CT}$ [$\Omega \cdot cm^2$] | $Y_{DL}$ [$\mu S \cdot cm^{-2} \cdot s^{nDL}$] | $n_{DL}$ | $R_{ads}$ [$\Omega \cdot cm^2$] | $Y_{ads}$ [$\mu S \cdot cm^{-2} \cdot s^{nads}$] | $n_{ads}$ | $L_{ads}$ [$H \cdot cm^2$] | $X^2$ |
|---|---|---|---|---|---|---|---|---|---|
| CR | | | | | | | | | |
| 1 | 6.3 | 23.7 | 87 | 0.92 | 4.0 | 89 | 1 | 0.47 | $1.8 \cdot 10^{-4}$ |
| 4 | 6.0 | 20.9 | 128 | 0.92 | 2.5 | 162 | 1 | 0.14 | $1.3 \cdot 10^{-4}$ |
| 8 | 6.1 | 16.2 | 224 | 0.90 | 1.9 | 269 | 1 | 0.11 | $2.5 \cdot 10^{-5}$ |
| 24 | 6.1 | 11.1 | 525 | 0.89 | 1.3 | 657 | 1 | 0.08 | $6.0 \cdot 10^{-5}$ |
| 550 °C | | | | | | | | | |
| 1 | 6.2 | 10.5 | 129 | 0.94 | 0.9 | 585 | 1 | 0.05 | $8.2 \cdot 10^{-5}$ |
| 4 | 6.1 | 9.5 | 250 | 0.92 | 1.1 | 391 | 1 | 0.05 | $6.5 \cdot 10^{-5}$ |
| 8 | 6.1 | 8.5 | 432 | 0.91 | 0.9 | 747 | 1 | 0.04 | $5.7 \cdot 10^{-5}$ |
| 24 | 6.2 | 7.3 | 1045 | 0.86 | 1.1 | 1010 | 1 | 0.05 | $6.9 \cdot 10^{-5}$ |
| 600 °C | | | | | | | | | |
| 1 | 6.6 | 7.1 | 264 | 0.89 | 0.6 | 561 | 1 | 0.02 | $6.5 \cdot 10^{-5}$ |
| 4 | 6.6 | 9.1 | 271 | 0.92 | 1.0 | 531 | 1 | 0.07 | $6.1 \cdot 10^{-5}$ |
| 8 | 6.6 | 8.5 | 425 | 0.92 | 0.8 | 929 | 1 | 0.04 | $4.3 \cdot 10^{-5}$ |
| 24 | 7.0 | 7.6 | 1101 | 0.89 | 1.0 | 1271 | 1 | 0.05 | $4.9 \cdot 10^{-5}$ |
| 700 °C | | | | | | | | | |
| 1 | 6.7 | 18.9 | 102 | 0.92 | 2.1 | 146 | 1 | 0.13 | $8.1 \cdot 10^{-5}$ |
| 4 | 6.5 | 17.3 | 165 | 0.92 | 1.9 | 261 | 1 | 0.10 | $7.9 \cdot 10^{-5}$ |
| 8 | 6.5 | 15.7 | 276 | 0.90 | 1.8 | 313 | 1 | 0.10 | $7.4 \cdot 10^{-5}$ |
| 24 | 6.4 | 12.6 | 631 | 0.90 | 1.5 | 638 | 1 | 0.08 | $4.7 \cdot 10^{-5}$ |

In Figure 10, SEM secondary electron (SE) top view images of all states after 1 h and 24 h of immersion in 0.1 M $H_2SO_4$ are depicted. The typical martensitic lancets are visible for the CR state after 1 h of immersion (Figure 10 (a)), while after 24 h immersion time small dimples are observed (Figure 10 (b)). Those dimples are also visible for the three ICA states ICA 550 °C (Figure 10 (d)), ICA 600 °C (Figure 10 (f)) and ICA 700 °C (Figure 10 (h)) after 24 h of immersion. In the top view images of ICA 550 °C (Figure 10 (c)), ICA 600 °C (Figure 10 (e)) and ICA 700 °C (Figure 10 (g)) after 1 h of immersion the reverted austenite and ferrite can be clearly identified due to the apparent selective dissolution of the reverted austenite and the consequent appearance of their morphologies.

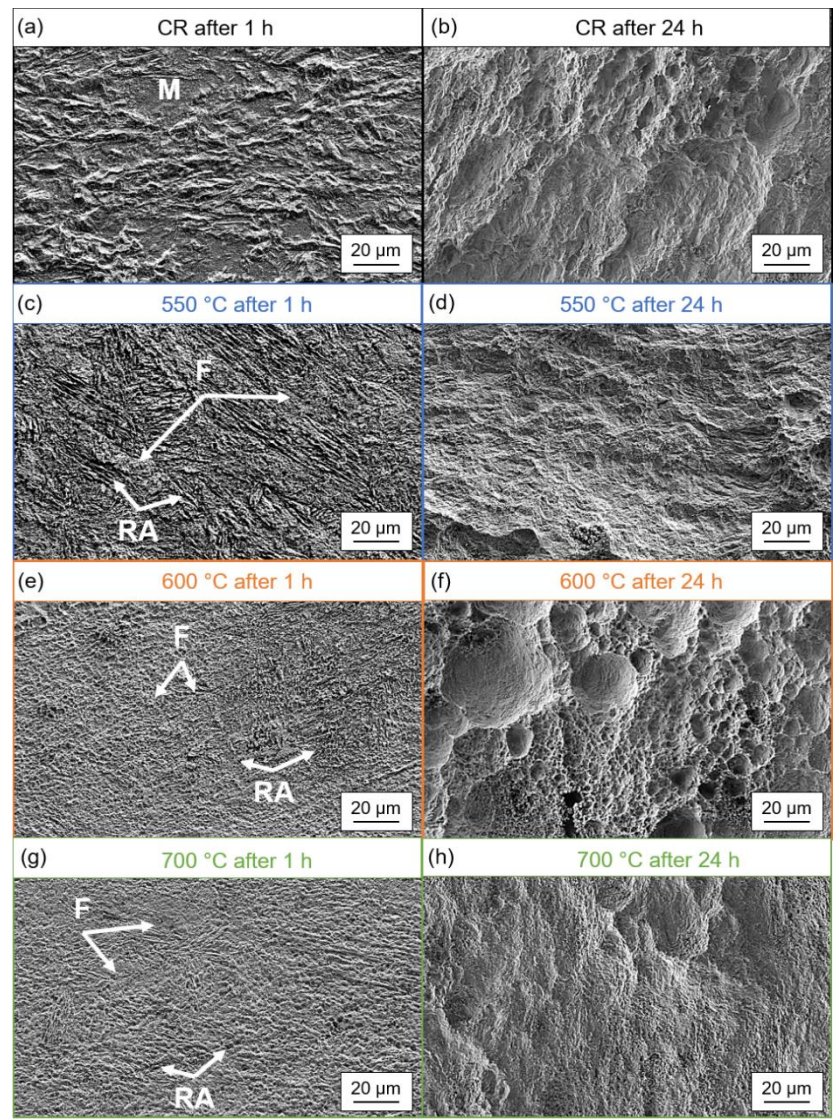

**Figure 10.** SEM secondary electron (SE) top view images after 1 h and 24 h of immersion in 0.1 M H₂SO₄ of CR (a and b), ICA 550 °C (c and d), ICA 600 °C (e and f), and ICA 700 °C (g and h).

In Figure 11, SEM backscattered electron (BSE) cross-section images of all states after immersion in 0.1 M H₂SO₄ for 1 h and 24 h are depicted. The CR state shows a relative uniform material removal after 1 h of immersion (Figure 11 (a)), while a preferred dissolution along the martensite lancets may be indicated after 24 h of immersion in Figure 11 (b). The cross-section images of ICA 550 °C after 1 h shows a selective dissolution of the reverted austenite (Figure 11 (c)), which is more pronounced after an immersion time of 24 h (Figure 11 (d)). The selective dissolution of the reverted austenite in ICA 600 °C is enhanced after 1 h (Figure 11 (e)) and 24 h (Figure 11 (f)) compared to the selective dissolution in ICA 550 °C. The deep dimples, which are already recognizable in the top view images of ICA 600 °C after an immersion time of 24 h, are also visible in the cross-section images due to the increased amount of selectively dissolved reverted austenite. The phenomenon of the selective corrosion attack along the reverted austenite is strongly reduced in ICA 700 °C after 1 h (Figure 11 (g)) and 24 h (Figure 11 (h)) of immersion compared to ICA 550 °C and ICA 600 °C. It is also notable that the surface of ICA 700 °C is uniformly dissolved after 24 h of immersion and only shows small signs of dimples, which is consistent with the top view images of the same state and immersion duration in Figure 10 (h).

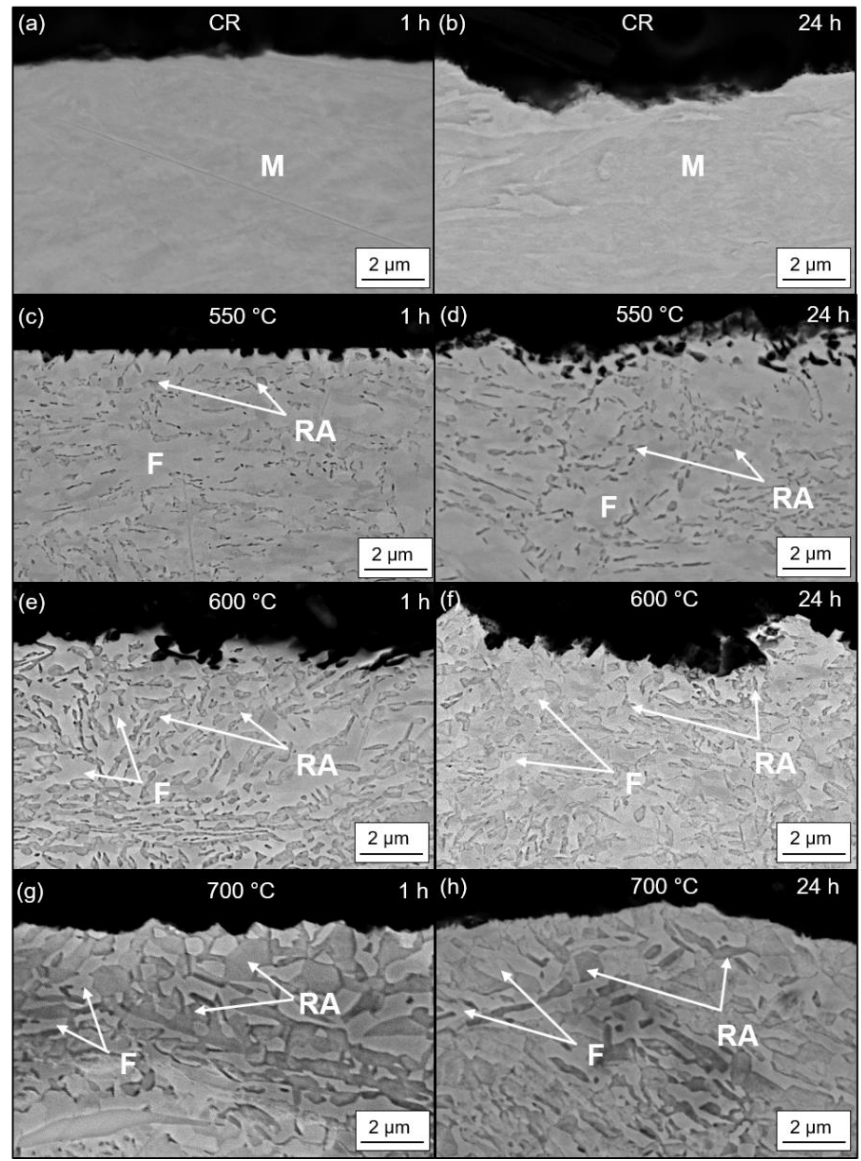

**Figure 11.** SEM backscattered electron (BSE) cross-section images after immersion of 1 h and 24 h in 0.1 M H₂SO₄ of CR (a and b), 550 °C (c and d), 600 °C (e and f), and 700 °C (g and h).

In Figure 12, the cumulative concentration c of Fe, Mn and Al in respect to the sample surface area A in the solution during immersion in 0.1 M H₂SO₄ is depicted. The ion concentration in the solution increases in all cases over time. After an immersion time of 1 h, the total ion concentration in the solution is the lowest for CR. The ion concentration is higher for ICA 550 °C and ICA 600 °C, while the ion concentration in the solution for ICA 700 °C lied in between. The same tendency is observed after 4 h. After an immersion time of 8 h, the ion concentrations in the solution for CR and ICA 700 °C overlap indicating a change in the dissolution processes. After 24 h of immersion, the ion concentration in the solution for ICA 700 °C is the lowest, while the ion concentration in the solution of CR is higher compared to ICA 700 °C and lower compared to ICA 550 °C and ICA 600 °C. The ion concentration values for the immersion tests of ICA 550 °C and ICA 600 °C are similar even after 24 h.

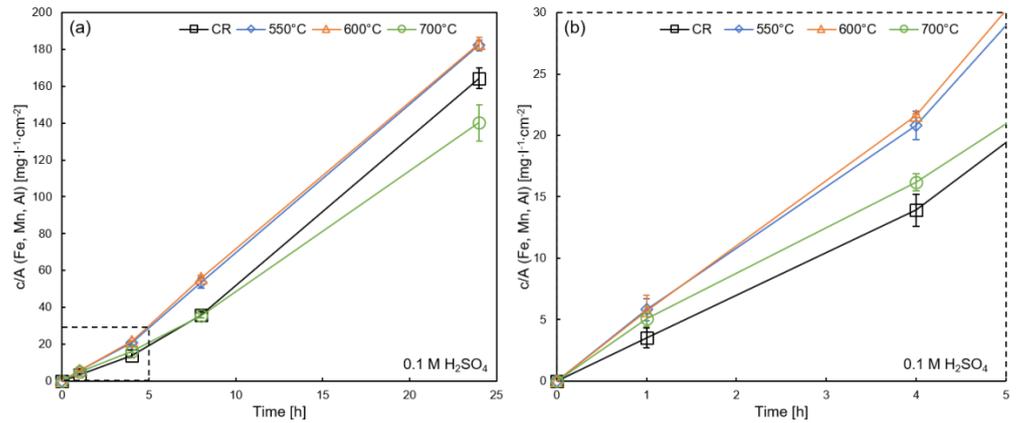

**Figure 12.** ICP-MS analysis of Fe, Mn and Al during immersion in 0.1 M $H_2SO_4$ (a) at all time steps and (b) at 1 and 4 h.

In Figure 13, the factor of dissolution ($F_d$) over the immersion time for all states in 0.1 M $H_2SO_4$ for the element Fe (Figure 13 (a)), Mn (Figure 13 (b)) and Al (Figure 13 (c)) is shown. The factor of dissolution is calculated using equation (1), which includes the mass fraction ($f_s$) in the solution and the mass fraction ($f_a$) in the alloy of the respective elements:

$$F_d = f_s / f_a \tag{1}$$

A $F_d$ value of more than 1 indicates a selective dissolution, while a value of less than 1 suggests a reduced mass loss of the respective element. A $F_d$ value of 1 means that the mass loss of an element is equal to its fraction in the alloy and therefore describes the equilibrium state. After 1 h of immersion, the factors of dissolution of all states show a selective dissolution of Fe, while Mn and Al show a lower mass loss compared to the respective fraction in the alloy. In case of CR, $F_d$ (Fe) and $F_d$ (Mn) stay constant after an immersion time of 4 h, while $F_d$ (Fe) decreases, and $F_d$ (Mn) increases slightly for ICA 700 °C. In case of ICA 550 °C and ICA 600 °C, $F_d$ (Fe) is decreased to a value of approximately 1 and $F_d$ (Mn) is increased above 1, which indicates a change in the selective dissolution behavior through the preferential dissolution of Mn. After 8 h of immersion, $F_d$ (Fe) approaches 1 for CR and ICA 700 °C, while $F_d$ (Mn) is increased above 1. On the contrary, $F_d$ (Mn) and $F_d$ (Fe) for ICA 550 °C and ICA 600 °C show a contrasting trend compared to CR and ICA 700 °C. After an immersion time of 24 h, the $F_d$ values of Fe, Mn and Al are equal for all states. While the $F_d$ (Fe) and $F_d$ (Mn) alternate, $F_d$ (Al) is decreased over time for all states and stays constantly below 1. After 1 h of immersion, $F_d$ (Al) for CR is slightly higher compared to $F_d$ (Al) for the other three states. While after 4 h of immersion $F_d$ (Al) are equal for all states, $F_d$ (Al) are higher for CR and ICA 700 °C compared to ICA 550 °C and ICA 600 °C.

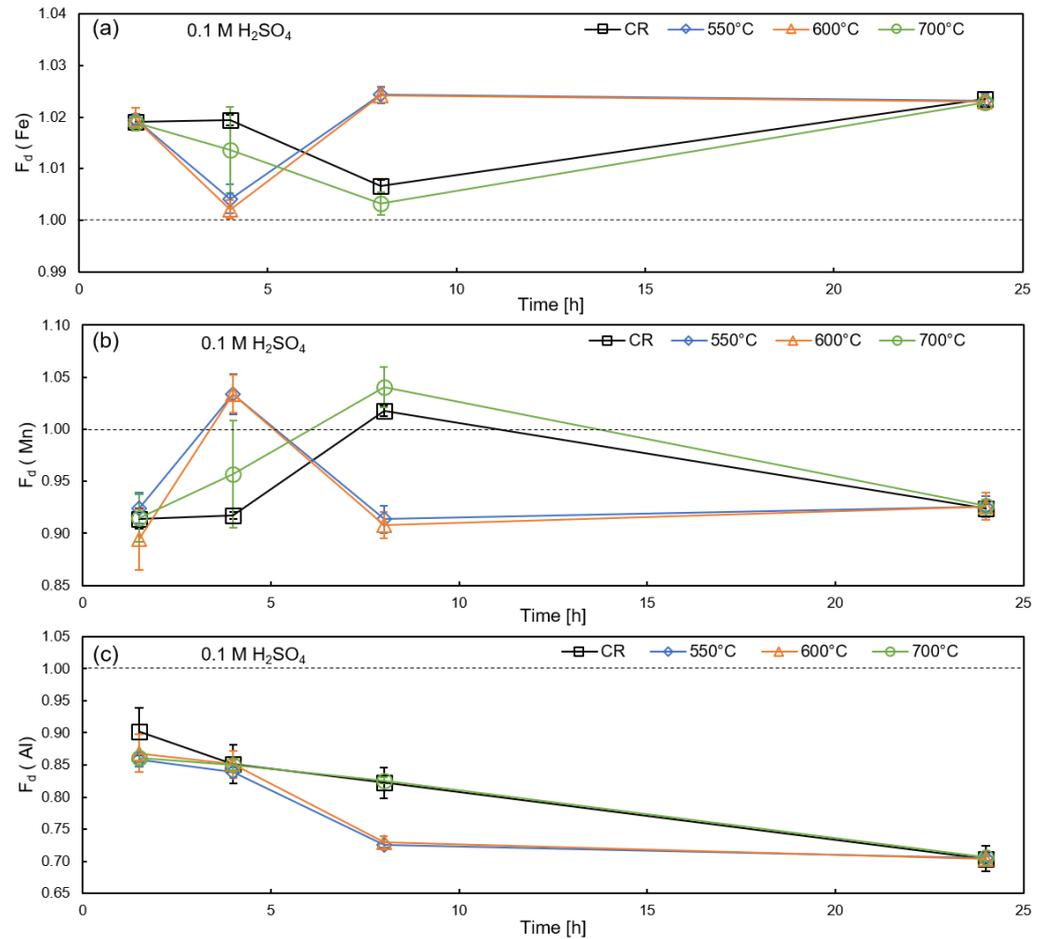

**Figure 13.** Factor of dissolution (F$_d$) of (a) Fe, (b) Mn and (c) Al for all states during immersion in 0.1 M H$_2$SO$_4$.

## 4. Discussion

### 4.1. Influences of ICA treatments on the microstructure evolution

The obtained microstructures of medium-Mn steel X6MnAl12-3 containing different amounts of reverted austenite were adjusted by applying ICA treatments at 550, 600 and 700 °C (between the critical temperatures A1 and A3). During these ICA treatments the martensitic microstructure (of the CR state) gradually transforms through Austenite-Reverted-Transformation (ART) [40] into dual-phase microstructures consisting of ferrite and austenite. The observed ultrafine grained microstructures are attributable to the high density of nucleation sites for ferrite transformation provided by the initial deformed martensitic microstructure, as the austenite preferentially nucleates at the packet boundaries as well as at the laths interfaces of martensite during ICA [7,41]. The growth of austenite during ICA treatments is believed to be controlled by the partitioning of austenite stabilizing elements (here C and Mn) as it was indicated from the compositional changes of ferrite and austenite in the ICA range based on the thermodynamic calculations represented in **Figure 3**. Such partitioning phenomena that locally occur at the ferrite/austenite interfaces of medium-Mn steel under investigation were previously depicted using atom probe tomography (APT) that enables identifying the nano-scale spatial chemical variations [42]. The XRD measurements and SEM micrographs demonstrated the increase in austenite fractions and the gradual vanishing of the austenite elongated grain morphology as the ICA temperature increases. The equiaxed and lamella (elongated) grain-types of morphologies were identified for medium-Mn steels of ferrite/austenite microstructures processed by ICA treatments [43].

The microstructural observations along with the XRD measurements are in a good agreement with phase evolution trend predicted by the equilibrium thermodynamic calculations listed in Table 4, although the absolute equilibrium fractions of austenite at the corresponding applied ICA temperatures are higher than the measured counterparts. The difference between the equilibrium austenite fractions and the measured ones is ascribed to the applied annealing time (2 h in the current study), which is practically too short to allow the equilibrium condition to be reached. In a previous study on the same material [3], energy intensive and longer-time (12 h) ICA treatments were applied, and relatively close-to-equilibrium austenite fractions were accordingly achieved in the temperature range from 555 to 650 °C. However, at 700 °C the final retained austenite fraction was higher than the prediction, which was explained by an ultrafine grain-induced austenite stability effect. Likewise, in the current study the equilibrium chemical compositions, particularly the decreasing Mn and C contents at the highest ICA temperatures (700 °C), imply a lower thermal stability of the reverted austenite, which was reflected by the estimated martensite start ($M_s$) temperature of 68 °C as well as with the predicted fraction of possible fresh martensite ($f_{\alpha}'$) of 37 % estimated by Koistinen-Marburger (KM) equation, according to the following equations [44]:

$$M_s = 547.58 - 596.914C - 28.389Mn + 8.827Al - 60.5V_{\gamma}{}^{-1/3} \tag{2}$$

$$f_{\alpha}' = 1 - \exp\,[-0.011\,(M_s - T)] \tag{3}$$

where C, Mn and Al are in wt.%, $V_{\gamma}$ is representing the volume of austenite grains (1 μm is used as the average grain diameter) and T is the temperature.

However, the XRD measurements and the microstructure investigations did not reveal the formation of fresh martensite emphasizing the influence of ultrafine grain size on the austenite stability. In addition, the observed variations in the Mn contents and austenite fractions driven by the different applied ICA temperatures have a profound effect on the corrosion behavior as will be deciphered in the following section. It was reported that the high dissolution rate of Mn leads to an inferior corrosion resistance of high-Mn steels [45].

**Table 4.** Austenite composition at different inter-critical annealing (ICA) temperatures based on the equilibrium thermodynamic calculations, and the predicted martensite start (Ms) temperature a

| ICA Temp. (°C) | Thermodynamic equilibrium calculations | | | | | Koistinen-Marburger-based calculations | | | Measured austenite (XRD) |
|---|---|---|---|---|---|---|---|---|---|
| | Austenite (vol.%) | Austenite composition (wt.%) | | | | Ms (°C) | fα' (vol.%) | Austenite (vol.%) | |
| | | C | Si | Mn | Al | | | | |
| 550 | 42 | 0.151 | 0.36 | 22.01 | 1.98 | -225 | 0 | 42 | 21.9 |
| 600 | 52 | 0.123 | 0.31 | 18.47 | 2.22 | -106 | 0 | 52 | 40.8 |
| 700 | 78 | 0.082 | 0.23 | 13.36 | 2.65 | 68 | 37 | 41 | 58.5 |

the fresh martensite ($f_{\alpha}'$) and austenite fractions.

### 4.2 Electrochemical testing and corrosion analysis

#### 4.2.1 Comparison with high-Mn steel and the role of Al

The corrosion current densities of all investigated states, which were determined by Tafel slope extrapolation, showed reduced values compared to a high-Mn steel containing 29.5 wt.% Mn, 3.1 wt.% Al and 1.4 wt.% Si [23] indicating the beneficial corrosion resistance of medium-Mn steel compared to high-Mn steel in acidic 0.1 M $H_2SO_4$. Those findings can be generally explained by the reduced Mn content of the X6MnAl12-3 material in the current investigation and the consequent reduction in Mn dissolution, because high Mn content was reported to highly contribute to the corrosion kinetics [23]. Al was mentioned to increase the passivity behavior of Fe-Mn-Al based steels in various solutions by several authors [20-22,27,46]. However, in the acidic solution 0.1 M $H_2SO_4$ with an approximate pH of $1.1 \pm 0.1$, Al preferentially dissolves as $Al^{3+}$ according to the Pourbaix diagram [47] and therefore 2.9 wt.% Al is not considered to contribute significantly to the formation of a stable passive film. That assumption is supported by the absence of passive region in the PDP curves, the low resistance in the EIS plots, and the absence of measurable corrosion products after immersion. Nevertheless, the formation of a thin layer during immersion at $E_{corr}$ is still expectable. An indication for the formation could be $F_d$ (Al) that was determined to be constantly below 1 for all states (Figure 13 (c)) and might be related to the incorporation of $Al^{3+}$ within the thin layer.

*4.2.2 Influence of ICA on corrosion behavior after 1 h in 0.1 M $H_2SO_4$*

The applied ICA treatments resulted in adjusting microstructures containing different amounts of reverted austenite with different content of alloying elements due to elements partitioning, in particular Mn, which consequently led to a divergent corrosion behavior in the investigations including PDP, EIS and ICP-MS. The shift of $E_{corr}$ to more negative potentials from CR to elevated ICA temperatures (Table 2) correlated with the increase in the reverted austenite fraction, indicating a more active behavior due to the increased austenite content. The chemical composition differences, especially the Mn content between austenite and ferrite caused a micro-galvanic coupling, which led to a selective dissolution of the more active austenite that acted as anode due to its relatively high Mn content. The selective dissolution of austenite for all ICA states is demonstrated in the SEM images of the corresponding cross-sections after immersion tests in Figure 11 (c-h). It is also clearly visible that the severity of the austenite dissolution is divergent for the different ICA states. Various factors can influence the micro-galvanic coupling of both phases, which need to be considered. The influencing factors involve the chemical content (mainly Mn) of both phases, the area of interfaces, and the element gradient at the interfaces.

Compared with ICA 600 °C, ICA 550 °C revealed a smaller $i_{corr}$ in PDP curves and a higher charge transfer resistance in EIS tests. The thermodynamically calculated Mn content in austenite (Table 4) is higher for ICA 550 °C compared to 600 °C, which consequently led to an increased potential difference between austenite and ferrite, thus an increased driving force for the selective austenite dissolution. However, with the ICA treatment temperature increasing from 550 °C to 600 °C, the austenite content increased and elongated austenite formed as shown in Figure 4 (g), exhibiting a higher area of austenite/ferrite interfaces in ICA 600 °C, which provided more corrosion sites. Despite of the higher driving force for micro-galvanic coupling between austenite and ferrite in ICA 550 °C, a relatively larger area of austenite/ferrite interfaces in ICA 600 °C contributed to the increased corrosion susceptibility. In case of ICA 700 °C, the thermodynamically calculated Mn content in austenite (13.36 wt.% Mn) approaches to the overall Mn content of the material (11.7 wt.% Mn), which decreased the driving force for micro-galvanic coupling and consequently the selective austenite dissolution compared to ICA 600 °C, which is clearly visible in the SEM cross-section image in Figure 11 (g). The reduced galvanic coupling of austenite and ferrite in duplex stainless steel by decreasing the Ni gradient between both phases was also previously reported to reduce the tendency of the preferential dissolution of the anodic phase [48]. Meanwhile, the increasing austenite fraction in ICA 700 °C resulted in an increasing anode-cathode surface area ratio, which in turn decreased the severity of austenite dissolution and corrosion kinetics. Besides, less austenite/ferrite interfaces were expected in ICA 700 °C than in ICA 600 °C, considering the microstructures shown in Figure 4 (c and d). Therefore, compared with ICA 600 °C, ICA 700 °C showed a decreased cumulative concentration of elements (Fe,

Mn and Al) in the solution, corrosion current density, and an increased charge transfer resistance.

### 4.2.3 Influence of ICA on corrosion behavior between 1 and 24 h in 0.1 M $H_2SO_4$

The immersion tests longer than 1 h indicated the ICA 700 °C state to stay the most corrosion resistant state compared to ICA 550 °C and ICA 600 °C by displaying the lowest cumulative concentration of Fe, Mn and Al and the highest charge transfer resistance. Moreover, the SEM images (Figure 11 (h)) of ICA 700 °C even showed that the dissolution of the surface remained uniform after 24 h immersion and no intensified selective dissolution of the austenite occurred. ICA 700 °C showed a higher corrosion resistance than CR after 8 h immersion with a lower cumulative concentration of elements in the solution (Figure 12) and a higher charge transfer resistance (Figure 9). The change might be related to other factors, such as dislocation defects in martensite, which were reported to be detrimental on the corrosion behavior in $H_2SO_4$ [49,50]. The immersion tests over a time range of 24 h indicated an assimilation of the corrosion behavior of the ICA 550 °C and ICA 600 °C states by the similar concentration of cumulative elements in the solution (Figure 12) and similar charge transfer resistances (Figure 9). Therefore, the micro-galvanic coupling seems to be most relevant in the early stage of the corrosion process during immersion, while other factors may dominate in the long-term immersion. It was reported that after the dissolution of the anodic phase, the detachment of the cathodic phase occurred for duplex stainless steel during immersion tests in acidic solution [51]. The dimples that were found in the SEM images (Figure 10) could be related to a detachment of the ferrite when the austenite was selectively dissolved.

### 4.2.4 Dissolution behavior of Mn and Fe

In order to consider the respective dissolution of Fe and Mn, the factors of dissolution were calculated and plotted over time for the different states (Figure 13). It needs to be stated that $F_d$ (Mn) and $F_d$ (Fe) were always close to 1 with subtle variations. Nevertheless, considering the fact that the Mn-rich reverted austenite was selectively dissolved, a preferred dissolution of Mn is expected after 1 h immersion, which was not the case in the current results. In contrast, $F_d$ (Fe) was constantly above 1, $F_d$ (Mn) only showed a value higher than 1 after 4 h for ICA 550 °C and ICA 600 °C as well as after 8 h for CR and ICA 700 °C. Considering the fact that the selective dissolution of the Mn-rich austenite was more pronounced for ICA 550 °C and ICA 600 °C, the presence of a $F_d$ (Mn) value above 1 was found earlier compared to CR and ICA 700 °C. The occurrence of the preferred dissolution of Mn after longer times might be explained by a time shift, because the total element concentration in the solution was determined after several time steps and small changes consequently might be measurable only after a certain time. Since the factors of dissolution of Fe and Mn were mostly above 1 and below 1, respectively, the selective dissolution of the austenite might not be the only determining factor, especially after longer immersion times. After the austenite was selectively dissolved in the early stage due to the micro-galvanic coupling of austenite and ferrite, the distance between the cathode (ferrite) and anode (austenite) increased, which reduced the driving force for the coupling. Therefore, the dissolution of ferrite, which contained lower amounts of Mn compared to austenite, will be more pronounced after a certain time. Due to the larger chemical gradient of Mn between austenite and ferrite and the larger cathode/anode ratio in ICA 550 °C compared to ICA 600 °C, the takeover could occur earlier for ICA 550 °C.

The factors of dissolution in Figure 13 were related to the overall composition of the material shown in Table 1. Since the phases differed mainly in the Mn content, the factors of dissolution were recalculated based on the thermodynamically determined Mn content in the austenite from Table 4 and are shown in Figure 14. The CR state was included as reference. The lowest $F_d$ (Mn) values were determined for ICA 550 °C and the highest for ICA 700 °C showing a contradicting trend with respect to the supposed dissolution tendency of the austenite. Considering the austenite fraction (Figure 5 (b)), which was the highest at ICA 700 °C and the lowest at ICA 550 °C, the highest austenite fraction appeared to be responsible for the highest dissolution of Mn. Furthermore, $F_d$ (Mn) remained constantly below 1, which indicated a reduced mass loss of Mn from the austenite.

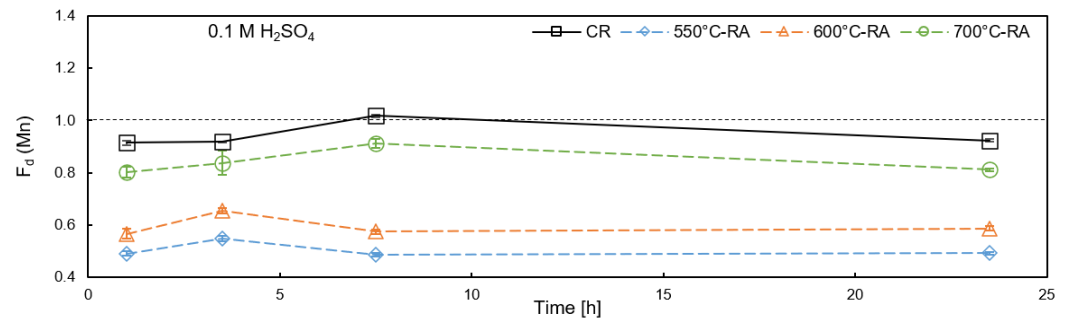

**Figure 14.** Factor of dissolution ($F_d$) of Mn based on the thermodynamically calculated Mn content in the reverted austenite (RA) for ICA 550 °C, ICA 600 °C and ICA 700 °C. The CR state is included as reference.

### 4.2.5 Initiation at the interfaces

The initiation of the selective dissolution occurred most likely at the austenite/ferrite interface. Bleck et al. [42] found a local Mn segregation at the austenite/ferrite interface of intercritically annealed (700 °C) medium-Mn steel, which is similar in the chemical composition compared to the X6MnAl12-3 material in the current investigation. The Mn segregation leads to a bilateral potential gradient, which is the highest in direction to the ferrite, since the Mn gradient is more pronounced. However, the Mn segregation at the interface makes it the most active region, which is therefore more likely to be dissolved first. It was stated by several authors [52-54] that the dissolution in duplex stainless steels in acidic solutions is also able to be initiated at the interfaces of e.g. intermetallic phases [54] or austenite/ferrite phase boundaries [52,53]. Sathirachinda et al. [54] mentioned that the selective dissolution at those interfaces could correlate to local chemical gradients and the consequent potential difference compared to the adjacent phases.

## 5. Conclusions

In this study, the influence of intercritical annealing (ICA) temperature on the electrochemical corrosion behavior of the medium-Mn steel X6MnAl12-3 in 0.1 M $H_2SO_4$ was investigated by PDP, EIS and ICP-MS. SEM investigations after 1 h and 24 h of immersion were conducted to visualized the corrosion damage. The study focused on the micro-galvanic coupling between reverted austenite and ferrite in dependence of the thermodynamically determined Mn partitioning. The following conclusions are drawn:

- The ICA treatments at 550, 600 and 700 °C led to a transformation from cold-rolled martensite to a ferrite/austenite (reverted) microstructure with increasing volume fraction of austenite and a decrease of Mn content within the austenite.
- The Mn partitioning led to a micro-galvanic coupling between ferrite (cathode) and reverted austenite (anode). The decreasing Mn gradient with increasing ICA temperature, and accordingly the fraction of reverted austenite, reduced the driving force for the micro-galvanic coupling. Apart from the Mn gradient, an increased interface area between austenite and ferrite enhanced the dissolution tendency of the anodic austenite.
- The immersion experiments between 1 h and 24 h indicated the decreasing influence of selective austenite dissolution and the transition to ferrite dissolution when the distance between ferrite and austenite is increased due to the selective dissolution of austenite in the early stage.

- The increase of the corrosion current density and decrease of the charge transfer resistance of ICA 600 °C compared to ICA 550 °C were related to the increase of interface area between austenite and ferrite due to the higher fraction of austenite. Those had a bigger influence on the micro-galvanic coupling than the chemical gradient of Mn.

- In case of ICA 700 °C, the chemical gradient of Mn was the smallest between austenite and ferrite, which changed the corrosion manner from local selective dissolution of reverted austenite to a uniform corrosion attack. Consequently, the corrosion current density and charge transfer resistance was reduced.

The control and design of Mn partitioning in reverted austenite during ICA of medium-Mn steels are of a prime importance for tailoring their mechanical properties. The current study additionally emphasizes its crucial role in controlling the corrosion resistance. This opens new opportunities to design new alloying concepts for medium-Mn steels that allow adjusting the characteristics of reverted austenite to optimize the corrosion properties with maintaining their promising mechanical properties. To that end, reducing the chemical gradient of Mn and the interface area between the reverted austenite and ferrite allow the adjustment of the severity of micro-galvanic coupling and consequently the selective dissolution of the austenite. The initiation of dissolution at the austenite/ferrite interfaces due to local Mn segregation in dependency of ICA temperatures and high-resolution surface analysis of potentially formed thin layers with incorporated Al are of special interest for future studies.

**Author Contributions:** Conceptualization, R.D.P. and C.H.; methodology, R.D.P., T.A., J.W. and J.N.; validation, R.D.P., T.A., J.W. and J.N.; formal analysis, R.D.P., T.A., J.W. and J.N.; investigation, R.D.P., T.A., J.W., J.N. and S.S.; resources, D.Z. and U.K.; data curation, R.D.P., T.A., J.W. and J.N.; writing—original draft preparation, R.D.P., T.A., J.W. and J.N.; writing—review and editing, R.D.P., T.A., J.W., J.N., C.H., S.S., U.K. and D.Z.; visualization, R.D.P., T.A., J.W., and J.N.; supervision, D.Z. and C.H.; All authors have read and agreed to the published version of the manuscript.

**Funding:** This research received no external funding.

**Acknowledgments:** The authors gratefully thank Katharina Utens (Chair of Corrosion and Corrosion Protection) for technical support.

**Conflicts of Interest:** The authors declare no conflict of interest.